\documentclass[twocolumn,showpacs,preprintnumbers,amsmath,amssymb]{revtex4}

\usepackage{graphicx}
\usepackage{dcolumn}
\usepackage{bm}

\newcommand{\bq}{\begin{equation}}
\newcommand{\eq}{\end{equation}}
\newcommand{\bqa}{\begin{eqnarray}}
\newcommand{\eqa}{\end{eqnarray}}
\newcommand{\nn}{\nonumber \\}

\def\be     {\begin{equation}}
\def\ee     {\end{equation}}
\def\bea        {\begin{eqnarray}}
\def\eea        {\end{eqnarray}}
\def\bnn    {\begin{eqnarray*}}
\def\enn    {\end{eqnarray*}}

\begin{document}

\title{Preformed heavy-electrons at the Quantum Critical  Point in heavy fermion compounds}
\author{Minh-Tien Tran$^{1,2}$, A. Benlagra$^3$, C. P\'epin$^{4}$ and Ki-Seok Kim$^{1,5}$}
\affiliation{$^1$Asia Pacific Center for Theoretical Physics,
POSTECH, Pohang, Gyeongbuk 790-784, Republic of Korea \\
$^2$Institute of Physics, Vietnamese Academy of Science and
Technology, P.O.Box 429, 10000 Hanoi, Vietnam \\ $^3$Institut
f$\ddot{u}$r Theoretische Physik, Technische Universit\"at Dresden, 01062 Dresden, Germany \\
$^4$Institut de Physique Th$\acute{e}$orique, CEA, IPhT, CNRS, URA
2306, F-91191 Gif-sur-Yvette, France \\ $^5$Department of Physics,
Pohang University of Science and Technology, Pohang, Gyeongbuk
790-784, Korea}

\date{\today}

\begin{abstract}
The existence of multiple energy scales is regarded as a signature of the
Kondo breakdown mechanism for explaining the quantum critical behavior of
certain heavy fermion compounds, like
YbRh$_{2}$Si$_{2}$.
The nature of the intermediate state between the heavy Fermi liquid and the quantum critical region, however,
remains elusive.  In this study we suggest an incoherent
heavy-fermion scenario, where inelastic scattering with novel soft
modes of the dynamical exponent $z = 3$ gives rise to non-Fermi
liquid physics for thermodynamics and transport despite the
formation of the heavy-fermion band. We discuss a crossover from
$z = 3$ to $z = 1$ for quantum phase fluctuations.
\end{abstract}

%\pacs{71.10.Hf, 71.30.+h, 71.10.-w, 71.10.Fd}

\maketitle

Research on quantum criticality has been one driving force in
modern condensed matter physics, where the universal scaling
reflects the non-perturbative nature of strong correlations
\cite{Review_QPT1,Review_QPT2}. The  observation of a regime with $T$-linear resistivity
%
%with temperature $T$
%
is the hallmark of quantum criticality in heavy fermions \cite{T_resistivity}.
%and the
%$\omega/T$ scaling of correlation functions
%
%with frequency $\omega$
 This observation  combined with the presence of anomalous exponents
calls for an interacting nature of the non-Fermi liquid fixed
point \cite{INS_Local_AF}. Heavy-fermion quantum criticality has
been regarded as a rule model, where competition between the Kondo
effect and Ruderman-Kittel-Kasuya-Yosida (RKKY) interaction gives
rise to a quantum phase transition from an antiferromagnetic metal
to a heavy-fermion Fermi-liquid. Non-Fermi liquid physics is
displayed in the quantum critical region
\cite{HF_Review1,HF_Review2}.

Two
competing theoretical frameworks have emerged, referred to as the Kondo breakdown
(KB) mechanism \cite{KB_Senthil,KB_Indranil,KB_Pepin,KB_Si} and
the spin-density-wave (SDW) scenario \cite{HMM,Rosch},
respectively.
%
%critical are Fermi-surface fluctuations associated with breakdown
%of the Kondo effect, or the spin-density-wave (SDW) scenario
%\cite{HMM,Rosch}, where antiferromagnetic spin fluctuations are
%critical modes.
%
Although these competing scenarios cover the $T$-linear transport
\cite{Rosch,KB_Indranil,KB_Pepin}, only the KB mechanism could
explain the divergent Gr\"uneisen ratio with the special critical
exponent of 2/3  in YbRh$_{2}$Si$_{2}$ \cite{Kim_GR}. In addition, another
KB scenario based on the slave-fermion representation uncovered
the diverging uniform spin susceptibility with an exponent $2/3$,
consistent with an experiment for YbRh$_{2}$Si$_{2}$
\cite{Kim_SF}.
%This is remarkable because critical ferromagnetic
%fluctuations appear in the antiferromagnetic quantum critical
%point (QCP) of the slave-fermion KB theory, which cannot be
%allowed in the SDW scenario.
%However, both theories fail to
%explain the $\omega/T$ scaling of the spin susceptibility. Only
%the local quantum critical scenario \cite{KB_Si}, regarded as
%another KB mechanism, turns out to work for such a scaling
%although it was observed only in CeCu$_{6-x}$Au$_{x}$
%\cite{INS_Local_AF}.

As shown in the above discussion, critical exponents can be
thought as a fingerprint of each scenario. These exponents can be
found from the Eliashberg approximation, where self-energy
corrections for both electrons and critical fluctuations are
introduced self-consistently
% giving rise to the scaling
%expression of the singular free energy
\cite{FM_QCP,Kim_LW}. However, the stability of the Eliashberg
framework has been questioned recently because electrons turn out
to be strongly interacting at quantum critical points (QCPs) even
in the large-$N$ limit, the cornerstone of the Eliashberg theory,
where $N$ is the number of fermion colors \cite{SSL,Max}. In this
respect it is desirable to find non-perturbative features beyond
the Eliashberg approximation.

Recently, two of us predicted violation of the Wiedemann-Franz law
at the KB QCP, where the existence of additional entropy carriers,
identified with charge-neutral spinon excitations, gives rise to
additive contributions for the thermal conductivity, resulting in
enhancement of the Lorentz number \cite{Kim_TR}. Furthermore, the
KB theory was claimed to show an abrupt collapse in the Seebeck
coefficient from the KB QCP to the SDW or spin liquid phase
because breakdown of the Kondo effect prohibits spinons from
carrying electric currents below a characteristic energy scale
$E^{*}$, where Fermi-surface fluctuations start to be frozen and
electrons in the f-orbital become localized \cite{Seebeck_Kim}.
These two features are based on reconstruction of the Fermi surface at the QCP,
distinguishing the KB scenario from the SDW
theory undebatably.

In this letter we investigate another signature of the KB
mechanism. The Hall coefficient has revealed a novel energy scale
$T^{*}$ higher than the Fermi-liquid temperature $T_{FL}$,
observed in the heavy-fermion side \cite{Hall1,Hall2}. It seems to
show an abrupt decrease at $T^{*}$, but the non-Fermi liquid
transport and thermodynamics are still observed in $T_{FL} < T <
T^{*}$. The abrupt change of the Hall coefficient is
believed to originate from the Fermi-surface reconstruction,  and has been corroborated by
observations of  a change in the  magnetoresistance  and  field dependence
of the magnetization \cite{brando}.

%thus
%regarded as the hall mark of the KB mechanism, the nature of this
%intermediate region is still elusive.

%\begin{figure}[t]
%\includegraphics[width=0.48\textwidth]{Phase_Diagram.eps}
%\caption{A schematic phase diagram in the preformed heavy-fermion
%(HF) scenario. $T^{*}$ corresponds to the mean-field transition
%temperature ($\rho_{b} \not= 0$) in the Kondo breakdown theory
%while the Fermi-liquid (FL) temperature $T_{FL}$ is reduced due to
%quantum phase fluctuations ($e^{i\theta_{b}}$) of the
%hybridization order parameter ($b = \sqrt{\rho_{b}}
%e^{i\theta_{b}}$). The intermediate region is identified with an
%incoherent heavy-fermion band, characterized by $\rho_{b} \not= 0$
%and $\langle e^{i\theta_{b}} \rangle = 0$. $T_{SL}$ denotes the
%spin-liquid temperature, and $E^{*}$ is the characteristic energy
%scale of the Kondo breakdown scenario, identified from the Seebeck
%coefficient \cite{Seebeck_Kim}. } \label{fig1}
%\end{figure}

Introducing the phase variable of the hybridization order
parameter into the KB theory, we propose that the intermediate
region is characterized by an incoherent heavy-fermion band, where
quantum phase fluctuations give rise to incoherent scattering of
heavy electrons and do not allow their Fermi liquid behaviors.
This preformed heavy-fermion scenario shows similarities
 with the preformed pair scenario for the pseudogap
phase of high $T_{c}$ cuprates \cite{Preformed_Pair}.

We start to discuss the Kondo effect in the single impurity
problem. As well known, the slave-boson mean-field theory allows for
a strong coupling fixed point, identified with the local Fermi
liquid state \cite{Hewson_Book}. However, it causes an artificial
second order transition at finite temperatures, which should not
exist in the single impurity problem. Fluctuation corrections are
introduced to check the stability of the mean-field state, where
they can be identified with contributions from vertex corrections
to the boson condensation \cite{Read_KI}. Such soft modes cause
the infrared $log$-divergence, argued to make condensation
prohibited, where an infinite-order summation based on the parquet
approximation will turn the $log$-divergence into a power-law
behavior. On the other hand, this treatment turns out to recover
correlation functions such as the specific heat coefficient and
spin susceptibility of the local Fermi liquid.
% where divergences
%of self-energy corrections are cancelled by those of vertex
%corrections due to the gauge invariance.

We apply this scheme to the heavy-fermion problem, described by an
effective Anderson lattice model \bqa && L = \sum_{i}
c_{i\sigma}^{\dagger}(\partial_{\tau} - \mu)c_{i\sigma} - t
\sum_{\langle ij \rangle} (c_{i\sigma}^{\dagger}c_{j\sigma} +
H.c.) \nn && + V \sum_{i} (d_{i\sigma}^{\dagger}c_{i\sigma} +
H.c.) + \sum_{i}d_{i\sigma}^{\dagger}(\partial_{\tau} +
\epsilon_{f})d_{i\sigma} \nn && + J \sum_{\langle ij \rangle}
\vec{S}_{i}\cdot\vec{S}_{j} , \eqa which shows competition between
the Kondo effect ($V$) and the RKKY interaction ($J$).
$c_{i\sigma}$ represents an electron in the conduction band with
its chemical potential $\mu$ and hopping integral $t$.
$d_{i\sigma}$ denotes an electron in the localized orbital with an
energy level $\epsilon_{f}$. The localized orbital experiences
strong repulsive interactions, thus either spin-$\uparrow$ or
spin-$\downarrow$ electrons can be occupied at most. This
constraint is incorporated in the U(1) slave-boson representation,
where the localized electron is decomposed into  holon and
spinon, $d_{i\sigma} = b_{i}^{\dagger} f_{i\sigma}$, supported by
the single-occupancy constraint $b_{i}^{\dagger}b_{i} +
f_{i\sigma}^{\dagger} f_{i\sigma} = S N$ in order to preserve the
physical space. $S = 1/2$ is the size of spin and $N$ is the spin
degeneracy, where the physical case is $N = 2$.

Rewriting the Anderson lattice model in terms of holons and
spinons, we obtain \bqa && Z = \int D c_{i\sigma} D f_{i\sigma} D
b_{i} D \chi_{ij} D \lambda_{i} e^{-\int_{0}^{\beta} d \tau L} ,
\nn && L = \sum_{i} c_{i\sigma}^{\dagger}(\partial_{\tau} -
\mu)c_{i\sigma} - t \sum_{\langle ij \rangle}
(c_{i\sigma}^{\dagger}c_{j\sigma} + H.c.) \nn && +
\frac{V}{\sqrt{N}} \sum_{i} (b_{i}f_{i\sigma}^{\dagger}c_{i\sigma}
+ H.c.) + \sum_{i}b_{i}^{\dagger} \partial_{\tau} b_{i} \nn && +
\sum_{i}f_{i\sigma}^{\dagger}(\partial_{\tau} +
\epsilon_{f})f_{i\sigma} - J \sum_{\langle ij \rangle} (
f_{i\sigma}^{\dagger}\chi_{ij}f_{j\sigma} + H.c.) \nn && + i
\sum_{i} \lambda_{i} (b_{i}^{\dagger}b_{i} + f_{i\sigma}^{\dagger}
f_{i\sigma} - S N) + N J \sum_{\langle ij \rangle} |\chi_{ij}|^{2}
, \eqa where the RKKY spin-exchange term for the localized orbital
is decomposed with the single occupancy constraint via exchange
hopping processes of spinons with a hopping parameter $\chi_{ij}$,
and $\lambda_{i}$ is a Lagrange multiplier field to impose the
single-occupancy constraint \cite{Supplementary}.

The saddle-point analysis with $b_{i} \rightarrow b$, $\chi_{ij}
\rightarrow \chi$, and $i\lambda_{i} \rightarrow \lambda$ reveals
a breakdown of the Kondo effect, where a spin-liquid Mott insulator
($b = 0$) arises with a small area of the Fermi surface in $J >
T_{K}$ while a heavy Fermi liquid ($b \not= 0$) obtains with a
large Fermi surface in $T_{K} > J$
\cite{KB_Senthil,KB_Indranil,KB_Pepin}. Here, $T_{K} = D
\exp\Bigl(\frac{\epsilon_{f}}{N \rho_{c}V^{2}}\Bigr)$ is the
single-ion Kondo temperature, where $\rho_{c} \approx (2D)^{-1}$
is the density of states for conduction electrons with the half
bandwidth $D$. Reconstruction of the Fermi surface occurs at $J
\simeq T_{K}$.

Quantum critical physics is characterized by critical fluctuations
of the hybridization order parameter, introduced in the Eliashberg
theory \cite{One_Loop_RG}.
%
%where self-energy corrections of electrons, spinons, and holons
%are taken into account fully self-consistently but vertex
%corrections are not incorporated.
%
Dynamics of critical Kondo fluctuations is described by $z = 3$
critical theory due to Landau damping of electron-spinon
polarization above an intrinsic energy scale $E^{*}$, while by $z
= 2$ dilute Bose gas model below $E^{*}$
\cite{KB_Indranil,KB_Pepin}. Here, $z$ is the dynamical critical
exponent, which tells the dispersion of bosonic modes. The energy
scale $E^{*}$ originates from the mismatch of Fermi surfaces of
conduction electrons and spinons, one of the central aspects in
the KB scenario. Physically, one may understand that quantum
fluctuations of the Fermi-surface reconfiguration start to be
frozen at $T \approx E^{*}$, thus the conduction electron's Fermi
surface dynamically decouples from the spinon's one below $E^{*}$.
%
%This is the energy scale, where the Seebeck coefficient collapses
%abruptly in the antiferromagnetic side \cite{Seebeck_Kim}.
%

We point out that the mean-field transition from the $z = 3$
quantum critical region to the heavy-fermion phase is identified
with $T^{*}$ of the Hall coefficient \cite{Hall1,Hall2}, where
quantum phase fluctuations of the holon order parameter reduce the
Fermi liquid temperature much. Decomposing the hybridization order
parameter into its amplitude and phase, $b = \sqrt{\rho_{b}}
e^{i\theta_{b}}$, and performing the continuum approximation in
terms of low energy fluctuations, we reach the following
expression \bqa && {\cal L} = c_{\sigma}^{*}(\partial_{\tau} -
\mu_{c})c_{\sigma} + \frac{1}{2m_{c}}|(\partial_{\boldsymbol{r}} +
i A_{\boldsymbol{r}}) c_{\sigma}|^{2} \nn && + \frac{V}{\sqrt{N}}
\sqrt{\rho_{b}} (e^{-i\theta_{b}} c_{\sigma}^{*}f_{\sigma} + H.c.)
\nn && + f_{\sigma}^{*}(\partial_{\tau} - \mu_{c} + \epsilon_{f} +
\lambda - ia_{\tau})f_{\sigma} +
\frac{1}{2m_{f}}|(\partial_{\boldsymbol{r}} -
ia_{\boldsymbol{r}})f_{\sigma}|^{2} \nn && + i \rho_{b} (
\partial_{\tau} \theta_{b} - i \lambda - a_{\tau}) + \frac{1}{2u_{b}}
( \partial_{\tau} \theta_{b} - i \lambda - a_{\tau})^{2} \nn && +
\frac{\rho_{b}}{2m_{b}} (\partial_{\boldsymbol{r}} \theta_{b} -
a_{\boldsymbol{r}} - A_{\boldsymbol{r}})^{2} + \frac{1}{4g^{2}}
(\partial_{\mu} a_{\nu} - \partial_{\nu} a_{\mu})^{2} \nn && +
\frac{u_{b}}{2} \rho_{b}^{2} - \lambda S N . \eqa $m_{c} =
\frac{1}{2 t}$ is the band mass of conduction electrons, and
$A_{\boldsymbol{r}}$ is an electromagnetic field. $m_{f} =
\frac{1}{2J\chi}$ is the band mass of spinons, and $\lambda$ is
the mean-field value of the Lagrange multiplier field with its
fluctuation part $a_{\tau}$. $a_{\boldsymbol{r}}$ originates from
the angular part of the hopping parameter, $\chi_{ij} = \chi
e^{ia_{ij}}$, playing the role of the U(1) gauge field. $m_{b}
\approx \frac{2J\chi}{V^{2}/N}$ is the band mass of holons,
originating from the electron-spinon polarization function at high
energies. The low energy physics from such Fermi-surface
fluctuations is given by the Landau damping term in the holon
(phase-fluctuation) propagator [Eq. (7)]. $u_{b}$ is a coupling
constant for local interactions between holons, given by
$\frac{u_{b}}{2} |b|^{4}$ and phenomenologically introduced, and
the second-order time-derivative term with $u_{b}$ results from
integration of $\delta \rho_{b}$ with $\rho_{b} \rightarrow
\rho_{b} + \delta \rho_{b}$ \cite{Supplementary}. $g$ is the
gauge-matter coupling constant.

This effective field theory is reduced to the slave-boson
mean-field theory when phase fluctuations are neglected, where
$\rho_{b}$ is identified with $b^{2}$. Thus, the mean-field
transition temperature is identified with $T^{*}$ because a finite
value of $\rho_{b}$ generates the heavy-fermion band, the Hall
coefficient being reduced due to the Fermi-surface reconstruction.
If $e^{i\theta_{b}} \approx 1 + i \theta_{b}$ is performed in the
Kondo-interaction term of the single impurity problem and phase
fluctuations are integrated over up to the second order, we can
see that an additional $log$-divergence in the spinon self-energy
cancels the $log$-divergence in the holon condensation, allowing
the amplitude ($\rho_{b}$) of the holon condensation to be finite
\cite{Supplementary}. This means that the mechanism for
disappearance of the holon condensation lies in transverse (phase)
fluctuations \cite{Read_KI}. On the other hand, such Goldstone
modes turn out to be not harmful for ordering in the heavy-fermion
problem with three dimensions. As a result, the heavy-fermion
Fermi-liquid state is stable against gaussian fluctuations of
Goldstone bosons $\theta_{b}$. However, the stability is not
guaranteed any more if quantum phase fluctuations are taken into
account beyond the gaussian order. The non-linear $\sigma$ model
approach is convenient to describe interactions between phase
modes \cite{NLsM}, where the phase factor is replaced with a
complex variable $\psi$. This complex field should be constrained
with the unimodular condition, $- \mu_{\psi} (|\psi|^{2} - 1)$
introduced into the effective Lagrangian, where $\mu_{\psi}$ is an
effective chemical potential.

Rewriting the effective Lagrangian Eq. (3) in terms of $\psi$, and
introducing quantum corrections self-consistently in the
Luttinger-Ward functional approach \cite{Kim_LW}, we obtain
coupled equations for self-energy corrections of electrons,
spinons, phases, and gauge fields. Since vertex corrections are
not taken into account, these self-consistent equations are
essentially the same as those of the quantum critical regime in
the KB theory \cite{KB_Indranil,KB_Pepin}. A novel feature beyond
the previous consideration is to introduce an additional energy
scale $\mu_{\psi}$, describing coherence of the heavy-fermion
band. The formation of the heavy-fermion band is determined from
$\rho_{b}$, controlled by $\lambda$.

We derive self-consistent equations for three order parameters
from the Luttinger-Ward free energy functional
\cite{Supplementary},
%Although vertex corrections are not
%introduced for self-energy corrections, such contributions are
%taken into account self-consistently for order parameters,
%incorporated by derivatives of self-energies with respect to order
%parameters. Schwinger-Dyson equations for order parameters have
%quantum corrections although most vertex corrections turn out to
%be cancelled, justifying our derivation \cite{Supplementary}.
%These coupled equations are given by
\bqa && \lambda -
\frac{V^{2}}{2\pi^{2}} \frac{\mathcal{K}_{F}^{c
2}}{\mathcal{V}_{F}^{c}} = 2 \frac{T V^{2}}{|\mu_{eff}^{\psi}|}
\frac{1}{\beta} \sum_{i\omega} \int \frac{d^{3} k}{(2\pi)^{3}}
g_{c}(k,i\omega) G_{f} (k,i\omega) , ~~~ \\ && \rho_{b} +
\frac{N}{\beta} \sum_{i\omega} \int \frac{d^{3} k}{(2\pi)^{3}}
G_{f} (k,i\omega) = N S , \\ && 1 -
\frac{\rho_{b}}{|\mu_{eff}^{\psi}|} \Bigl\{ \lambda -
\frac{V^{2}}{2\pi^{2}} \frac{\mathcal{K}_{F}^{c
2}}{\mathcal{V}_{F}^{c}} \Bigr\} = - \frac{1}{\beta}
\sum_{i\Omega} \int \frac{d^{3} q}{(2\pi)^{3}} G_{\psi}
(q,i\Omega) , \eqa where $G_{f(\psi)}(k,i\omega)$ is the
renormalized Green's function of spinons (phases) with the
heavy-fermion band and $g_{c}(k,i\omega)$ is the bare Green's
function of electrons. $\mathcal{V}_{F}^{c(f)}$ and
$\mathcal{K}_{F}^{c(f)}$ are renormalized Fermi velocity and
renormalized Fermi momentum of electrons (spinons) in the
heavy-fermion band, respectively. $\mu_{eff}^{\psi} = \mu_{\psi} -
\rho_{b} \lambda - \Sigma_{\psi}(0,0)$ is an effective chemical
potential, which determines the Fermi-liquid temperature $T_{FL}$,
where the constant contribution of the $\psi$ self-energy is
$\Sigma_{\psi}(0,0) = - \frac{V^{2} \rho_{b}}{2\pi^{2}}
\frac{\mathcal{K}_{F}^{c 2}}{\mathcal{V}_{F}^{c}}$.

\begin{figure}[t]
\includegraphics[width=0.50\textwidth]{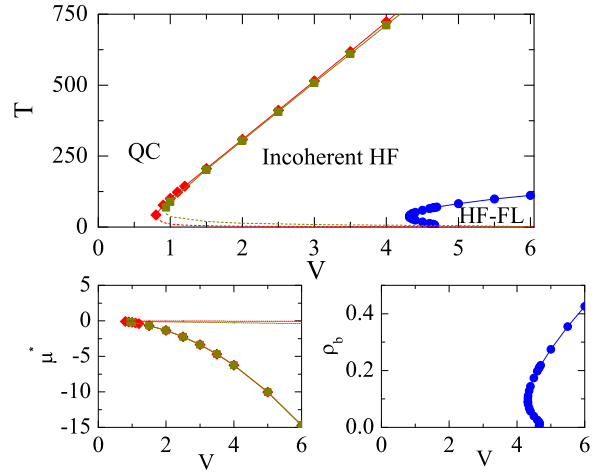}
\caption{A phase diagram in the preformed heavy-fermion scenario,
where QC, HF, and FL denote quantum critical, heavy fermion, and
Fermi liquid, respectively. $T^{*}(V)$ corresponds to the
mean-field transition temperature ($\rho_{b} = 0$) in the Kondo
breakdown theory while the Fermi-liquid temperature $T_{FL}(V)$ is
much reduced due to quantum phase fluctuations of the
hybridization order parameter ($\langle e^{i\theta_{b}} \rangle =
0$). Reentrant behaviors are found in both $T^{*}$ and $T_{FL}$
numerically, but it is not clear whether this effect is
fundamental or not due to quantum fluctuations. $\mu_{eff}^{\psi
*}(V,T^{*})$
%
%with $\lambda^{*}(V,T^{*})$
%
and $\rho_{b}(V,T_{FL})$
%
%with $\lambda_{FL}(V,T_{FL})$
%
are also shown, where $m_{c} = 0.01 m_{f}$, $\mu = m_{f}^{-1}$,
and $\epsilon_{f} = 10 m_{f}^{-1}$ are used with cutoffs of
$\Lambda_{q} = \Lambda_{\nu} = 10 m_{f}^{-1}$ for the red-diamond
and blue-circle lines and $\Lambda_{q} = \Lambda_{\nu} = 50
m_{f}^{-1}$ for the green-square line \cite{Supplementary}. The
unit of each axis is $m_{f}^{-1}$.} \label{fig1}
\end{figure}

We perform the numerical analysis, where self-energy corrections
are evaluated analytically. A detailed procedure can be found in
our supplementary material \cite{Supplementary}. Figure 1 displays
the intermediate state, where $\rho_{b}$ is finite, resulting in
the formation of the heavy-fermion band, while its coherence is
not achieved yet, reflected in the fact that the chemical
potential $- \mu_{eff}^{\psi} > 0$. $T^{*}$ is characterized by
$\rho_{b}(T^{*}) = 0$, and $T_{FL}$ is determined by
$\mu_{eff}^{\psi}(T_{FL}) = 0$.
%
%The right panel corresponds to the case when quantum corrections
%are neglected in Eqs. (4), (5), and (6). It is clear that quantum
%corrections reduce $T_{FL}$.
%
%Decreasing temperature, $\mu_{eff}$ vanishes to determine $T_{FL}$.

In the preformed heavy-fermion state the self-energy correction of
$\psi$ is governed by Landau damping from incoherent heavy
fermions, $\Delta \Sigma_{\psi}(q,i\Omega) = \gamma
\frac{|\Omega|}{q}$, where the damping coefficient is given by
$\gamma = V^{2} \rho_{b} \frac{\mathcal{K}_{F}^{f 2}}{4 \pi
\mathcal{V}_{F}^{f} \mathcal{V}_{F}^{c}}$. Then, the imaginary
part of the $\psi$ propagator becomes \bqa && - \Im
G_{\psi}(q,\Omega) = \frac{\gamma \Omega q}{\gamma^{2} \Omega^{2}
+ q^{2} \Bigl(\frac{\rho_{b}}{2m_{b}} q^{2} -
\mu_{eff}^{\psi}\Bigr)^{2}} . \eqa This expression displays a
crossover from $z = 3$ to $z = 1$ at $T_{1} \approx
\frac{1}{\gamma} \sqrt{\frac{2m_{b}}{\rho_{b}}} (-
\mu_{eff}^{\psi})^{\frac{3}{2}}$ as far as $T_{1}$ remains larger
than $T_{FL}$. In $T_{FL} < T < T_{1}$, it is given by $- \Im
G_{\psi}(q,\Omega) = \frac{\gamma \Omega q}{\gamma^{2} \Omega^{2}
+ \mu_{eff}^{\psi 2} q^{2}}$.

Inserting the $z = 1$ propagator into self-energy equations for
fermions, one finds that scattering with such fluctuations is
less relevant for self-energy corrections of fermions than
Fermi-liquid corrections in three dimensions. As a result, we
expect that the $T$-linear resistivity due to scattering with $z =
3$ critical modes \cite{KB_Indranil} becomes smoothly transformed into the
Fermi-liquid resistivity in the intermediate phase.

Recently, the $T^{*}$ line was proposed to be a Lifshitz
transition \cite{Vojta_Lifshitz}, motivated by the observation that
isoelectronic chemical doping does not change $T^{*}$ while it
affects  the N\'eel temperature seriously
\cite{brando}. On the other hand, non-isoelectronic
chemical doping changes $T^{*}$ clearly, when $Rh$ is replaced
with $Fe$ \cite{Chemical_Doping_II}. We believe that this issue
should be clarified.

In conclusion, we uncovered a new incoherent heavy-fermion state  which can be relevant to
the nature of the intermediate region of $T_{FL} < T < T^{*}$. The
mechanism turns out to be existence of quantum phase fluctuations
in the hybridization order parameter. Despite the formation of the
heavy-fermion band, this intermediate state will show non-Fermi
liquid physics in transport and thermodynamics due to scattering
with such $z = 3$ soft modes. The non-Fermi liquid physics become
transformed into the Fermi liquid physics continuously, as the
$z = 3$ critical mode turns into  $z = 1$, irrelevant for
fermion dynamics.

This work was supported by the National Research Foundation of
Korea (NRF) grant funded by the Korea government (MEST) (No.
2011-0074542). M.-T. was also supported by the Vietnamese
NAFOSTED.

\begin{widetext}

\appendix

\section{To construct the Luttinger-Ward functional}

Based on the nonlinear $\sigma$ model approach, we analyze an
effective Lagrangian Eq. (3). Introducing $e^{i\theta_{b}}
\rightarrow \psi$ with the unimodular constraint $|\psi|^{2} = 1$,
we rewrite Eq. (3) as follows \bqa && {\cal L} \approx
c_{\sigma}^{*}(\partial_{\tau} - \mu_{c})c_{\sigma} +
\frac{1}{2m_{c}}|(\partial_{i} + i A_{i}) c_{\sigma}|^{2} +
\frac{V}{\sqrt{N}} \sqrt{\rho_{b}} (\psi^{*}
c_{\sigma}^{*}f_{\sigma} + H.c.) \nn && +
f_{\sigma}^{*}(\partial_{\tau} - \mu_{c} + \epsilon_{f} + \lambda
- ia_{\tau})f_{\sigma} + \frac{1}{2m_{f}}|(\partial_{i} -
ia_{i})f_{\sigma}|^{2} \nn && + \rho_{b} \psi^{*} (\partial_{\tau}
+ \lambda - i a_{\tau}) \psi - \frac{1}{2u_{b}} [\psi^{*}
(\partial_{\tau} + \lambda - i a_{\tau}) \psi]^{2} +
\frac{\rho_{b}}{2m_{b}} |(\partial_{i} - i a_{i} - i A_{i})
\psi|^{2} - \mu_{\psi} (|\psi|^{2} - 1) \nn &&  + \frac{u_{b}}{2}
\rho_{b}^{2} - \lambda S N + \frac{1}{4g^{2}} (\partial_{\mu}
a_{\nu} - \partial_{\nu} a_{\mu})^{2} ,  \eqa where $\mu_{\psi}$
plays the role of an effective chemical potential, imposing the
rotor constraint. As a result, three order parameters appear to be
$\rho_{b}$, $\lambda$, and $\mu_{\psi}$ beyond the slave-boson
mean-field analysis. Introduction of $\mu_{\psi}$ gives rise to a
novel energy scale, determining the coherence of the heavy-fermion
band.

One can derive an effective action from our effective field theory
Eq. (A1), taking into account quantum corrections
self-consistently in the Eliashberg approximation, where
self-energy corrections are introduced, but vertex corrections are
neglected. The Eliashberg approximation results in the following
effective action \bqa && S = \int_{0}^{\beta} d \tau \int d^{3} r
\int_{0}^{\beta} d \tau' \int d^{3} r' \Bigl[
c_{\sigma}^{*}(r,\tau) \Bigl\{ \Bigl(
\partial_{\tau} - \mu_{c} - \frac{\partial_{i}^{2}}{2m_{c}} \Bigr)
\delta(\tau-\tau') \delta^{3}(r-r') + \Sigma_{c}(r-r',\tau-\tau')
\Bigr\} c_{\sigma}(r',\tau') \nn && - N
\Sigma_{c}(r-r',\tau-\tau') G_{c}(r'-r,\tau'-\tau) \Bigr] \nn && +
\int_{0}^{\beta} d \tau \int d^{3} r \int_{0}^{\beta} d \tau' \int
d^{3} r' \Bigl[ f_{\sigma}^{*}(r,\tau) \Bigl\{ \Bigl(
\partial_{\tau} - \mu_{c} + \epsilon_{f} + \lambda - \frac{\partial_{i}^{2}}{2m_{f}} \Bigr)
\delta(\tau-\tau') \delta^{3}(r-r') + \Sigma_{f}(r-r',\tau-\tau')
\Bigr\} f_{\sigma}(r',\tau') \nn && - N
\Sigma_{f}(r-r',\tau-\tau') G_{f}(r'-r,\tau'-\tau) \Bigr] \nn && +
\int_{0}^{\beta} d \tau \int d^{3} r \int_{0}^{\beta} d \tau' \int
d^{3} r' \Bigl[ \psi^{*}(r,\tau) \Bigl\{ \Bigl( [\rho_{b} -
\lambda/u_{b}] \partial_{\tau} - \frac{1}{2u_{b}}
\partial_{\tau}^{2} - \mu_{\psi} + \rho_{b} \lambda -
\frac{\rho_{b}}{2m_{b}}
\partial_{i}^{2} \Bigr) \delta(\tau-\tau') \delta^{3}(r-r') \nn &&
+ \Sigma_{\psi}(r-r',\tau-\tau') \Bigr\} \psi(r',\tau') +
\Sigma_{\psi}(r-r',\tau-\tau') G_{\psi}(r'-r,\tau'-\tau) \Bigr]
\nn && + \int_{0}^{\beta} d \tau \int d^{3} r \int_{0}^{\beta} d
\tau' \int d^{3} r' \Bigl[ \frac{1}{2} a_{\mu}(r,\tau) \Bigl\{
\Bigl( - \frac{\partial_{\tau}^{2} +
\partial_{i}^{2}}{g^{2}} \Bigr) P_{\mu\nu}^{T} \delta(\tau-\tau') \delta^{3}(r-r') +
\Pi_{\mu\nu}(r-r',\tau-\tau') \Bigr\} a_{\nu}(r',\tau') \nn && +
\Pi_{\mu\nu}(r-r',\tau-\tau') D_{\mu\nu}(r'-r,\tau'-\tau) \Bigr]
\nn && - V^{2}\rho_{b} \int_{0}^{\beta} d \tau \int d^{3} r
\int_{0}^{\beta} d \tau' \int d^{3} r' G_{\psi}(r-r',\tau-\tau')
G_{f}(r'-r,\tau'-\tau) G_{c}(r-r',\tau-\tau') \nn && - \frac{N}{2}
\int_{0}^{\beta} d \tau \int d^{3} r \int_{0}^{\beta} d \tau' \int
d^{3} r' v_{\mu}^{f} D_{\mu\nu} (r-r',\tau-\tau') v_{\nu}^{f}
G_{f}(r'-r,\tau'-\tau) G_{f}(r-r',\tau-\tau') \nn && - \frac{1}{2}
\int_{0}^{\beta} d \tau \int d^{3} r \int_{0}^{\beta} d \tau' \int
d^{3} r' v_{\mu}^{\psi} D_{\mu\nu} (r-r',\tau-\tau')
v_{\nu}^{\psi} G_{\psi}(r'-r,\tau'-\tau) G_{\psi}(r-r',\tau-\tau')
\nn && - \beta L^{3} \Bigl( \frac{\lambda^{2}}{2u_{b}} -
\mu_{\psi} - \frac{u_{b}}{2} \rho_{b}^{2} + N S \lambda \Bigr) ,
\eqa where $\Sigma_{c}(r-r',\tau-\tau')$,
$\Sigma_{f}(r-r',\tau-\tau')$, $\Sigma_{\psi}(r-r',\tau-\tau')$,
and $\Pi_{\mu\nu}(r-r',\tau-\tau')$ are self-energy corrections of
electrons, spinons, $\psi$, and gauge fields, respectively, and
$G_{c}(r-r',\tau-\tau')$, $G_{f}(r-r',\tau-\tau')$,
$G_{\psi}(r-r',\tau-\tau')$, and $D_{\mu\nu}(r-r',\tau-\tau')$ are
their Green's functions. Although vertex corrections are neglected
for self-energy calculations, such contributions are introduced
self-consistently into three coupled equations for order
parameters. A way how to derive this effective action is shown in
Ref. \cite{Kim_LW}.

Performing the Fourier transformation and integrating over all
field variables, we find the Luttinger-Ward functional \bqa &&
F_{LW}[\Sigma_{c}(k,i\omega),\Sigma_{f}(k,i\omega),\Sigma_{\psi}(q,i\Omega),\Pi_{ij}(q,i\Omega),\rho_{b},\lambda,\mu_{\psi}]
= - L^{3} \Bigl( \frac{\lambda^{2}}{2u_{b}} - \mu_{\psi} -
\frac{u_{b}}{2} \rho_{b}^{2} + N S \lambda \Bigr) \nn && -
\frac{N}{\beta} \sum_{i\omega} \int \frac{d^{3} k}{(2\pi)^{3}}
\Bigl\{ \ln \Bigl( - G_{c}^{-1}(k,i\omega) \Bigr) +
\Sigma_{c}(k,i\omega) G_{c}(k,i\omega) \Bigr\} \nn && -
\frac{N}{\beta} \sum_{i\omega} \int \frac{d^{3} k}{(2\pi)^{3}}
\Bigl\{ \ln \Bigl( - G_{f}^{-1}(k,i\omega) \Bigr) +
\Sigma_{f}(k,i\omega) G_{f}(k,i\omega) \Bigr\} \nn && +
\frac{1}{\beta} \sum_{i\Omega} \int \frac{d^{3} q}{(2\pi)^{3}}
\Bigl\{ \ln \Bigl( - G_{\psi}^{-1}(q,i\Omega) \Bigr) +
\Sigma_{\psi}(q,i\Omega) G_{\psi}(q,i\Omega) \Bigr\} \nn && +
\frac{1}{\beta} \sum_{i\Omega} \int \frac{d^{3} q}{(2\pi)^{3}}
\Bigl\{ \ln \Bigl( - D^{-1}(q,i\Omega) \Bigr) + \Pi(q,i\Omega)
D(q,i\Omega) \Bigr\} \nn && - V^{2}\rho_{b} \frac{1}{\beta}
\sum_{i\omega} \int \frac{d^{3} k}{(2\pi)^{3}} \frac{1}{\beta}
\sum_{i\Omega} \int \frac{d^{3} q}{(2\pi)^{3}} G_{\psi}(q,i\Omega)
G_{f}(k,i\omega) G_{c}(k+q,i\omega+i\Omega) \nn && - \frac{N}{2}
\frac{1}{\beta} \sum_{i\omega} \int \frac{d^{3} k}{(2\pi)^{3}}
\frac{1}{\beta} \sum_{i\Omega} \int \frac{d^{3} q}{(2\pi)^{3}}
F(k,q) D (q,i\Omega) G_{f}(k,i\omega) G_{f}(k+q,i\omega+i\Omega)
\nn && - \frac{\rho_{b}^{2} }{2} \frac{1}{\beta} \sum_{i\Omega}
\int \frac{d^{3} q}{(2\pi)^{3}} \frac{1}{\beta} \sum_{i\nu} \int
\frac{d^{3} l}{(2\pi)^{3}} B(q,l) D (l,i\nu) G_{\psi}(q,i\Omega)
G_{\psi}(q+l,i\Omega+i\nu) , \eqa where $G_{c}(k,i\omega)$,
$G_{f}(k,i\omega)$, $G_{\psi}(k,i\omega)$, and $D(q,i\Omega)$ are
Green's functions of electrons, spinons, phases, and gauge fields,
respectively, given by \bqa && G_{c}(k,i\omega) = \frac{1}{i\omega
+ \mu_{c} - \frac{k^{2}}{2m_{c}} - \Sigma_{c}(k,i\omega)} , ~~~~~
G_{f}(k,i\omega) = \frac{1}{i\omega + \mu_{c} - \epsilon_{f} -
\lambda  - \frac{k^{2}}{2m_{f}} - \Sigma_{f}(k,i\omega)} , \nn &&
G_{\psi}(q,i\Omega) = \frac{1}{(\rho_{b} - \lambda/u_{b})
(i\Omega) - \frac{\Omega^{2}}{2u_{b}} - \frac{\rho_{b}}{2m_{b}}
q^{2} + \mu_{\psi} - \rho_{b} \lambda - \Sigma_{\psi}(q,i\Omega)}
, \nn && D(q,i\Omega) = - \frac{1}{\frac{\Omega^{2} +
q^{2}}{g^{2}} + \Pi(q,i\Omega)} , ~~~~~ D_{ij}(q,i\Omega) =
D(q,i\Omega) P_{ij}^{T}(q) , ~~~~~ \Pi_{ij}(q,i\Omega) =
\Pi(q,i\Omega) P_{ij}^{T}(q) . \eqa $P_{ij}^{T}(q)$ is the
projection operator to the transverse component, and \bqa &&
F(k,q) = \frac{1}{2} \sum_{i,j = 1}^{2} v_{i}^{f} \Bigl(
\delta_{ij} - \frac{q_{i}q_{j}}{q^{2}} \Bigr) v_{j}^{f} , ~~~~~
v_{i}^{f} = \frac{k_{i} + q_{i}/2}{m_{f}} , \nn && B(q,l) =
\frac{1}{2} \sum_{i,j = 1}^{2} v_{i}^{\psi} \Bigl( \delta_{ij} -
\frac{l_{i}l_{j}}{l^{2}} \Bigr) v_{j}^{\psi} , ~~~~~ v_{i}^{\psi}
= \frac{q_{i} + l_{i}/2}{m_{b}} , \nonumber \eqa where $v_{i}^{f}$
and $v_{i}^{\psi}$ are velocities of spinons and phases in the
$i$-direction.

Minimizing the free energy functional with respect to all
self-energies, we obtain self-consistent Eliashberg equations \bqa
&& \Sigma_{c}(k,i\omega) = - \frac{V^{2} \rho_{b}}{N}
\frac{1}{\beta} \sum_{i\Omega} \int \frac{d^{3}
q}{(2\pi)^{3}}G_{\psi}(q,i\Omega) G_{f}(k - q,i\omega - i\Omega) ,
\nn && \Sigma_{f}(k,i\omega) = - \frac{V^{2} \rho_{b}}{N}
\frac{1}{\beta} \sum_{i\Omega} \int \frac{d^{3} q}{(2\pi)^{3}}
G_{\psi}(q,i\Omega) G_{c}(k + q,i\omega + i\Omega) -
\frac{1}{\beta} \sum_{i\Omega} \int \frac{d^{3} q}{(2\pi)^{3}}
F(k,q) D (q,i\Omega) G_{f}(k,i\omega) , \nn &&
\Sigma_{\psi}(q,i\Omega) = V^{2}\rho_{b} \frac{1}{\beta}
\sum_{i\omega} \int \frac{d^{3} k}{(2\pi)^{3}} G_{f}(k,i\omega)
G_{c}(k+q,i\omega+i\Omega) + \rho_{b}^{2} \frac{1}{\beta}
\sum_{i\nu} \int \frac{d^{3} l}{(2\pi)^{3}} B(q,l) D (l,i\nu)
G_{\psi}(q+l,i\Omega+i\nu) , \nn && \Pi(q,i\Omega) = \frac{N}{2}
\frac{1}{\beta} \sum_{i\omega} \int \frac{d^{3} k}{(2\pi)^{3}}
F(k,q) G_{f}(k,i\omega) G_{f}(k+q,i\omega+i\Omega)
+\frac{\rho_{b}^{2} }{2} \frac{1}{\beta} \sum_{i\nu}
\int\frac{d^{3} l}{(2\pi)^{3}} B(q,l) G_{\psi}(q,i\Omega)
G_{\psi}(q+l,i\Omega+i\nu) . \nn \eqa We note that these equations
are essentially the same as those in the quantum critical regime
of the Kondo breakdown theory \cite{KB_Indranil,KB_Pepin}, where
the self-energy and Green's function of $\psi$ are identified with
those of $b$.

\section{To evaluate self-energy corrections}

\subsection{Self-energy corrections for the heavy-fermion band}

In order to describe the heavy-fermion band without condensation
of $\psi$, we separate fermion self-energy corrections as follows
\bqa && \Sigma_{c}(k,i\omega) = \Phi_{c}(k,i\omega) + \Delta
\Sigma_{c}(i\omega) , ~~~~~ \Sigma_{f}(k,i\omega) =
\Phi_{f}(k,i\omega) + \Delta \Sigma_{f}(i\omega) , \eqa where
$\Phi_{c}(k,i\omega)$ and $\Phi_{f}(k,i\omega)$ are associated
with the formation of the heavy-fermion band, and $\Delta
\Sigma_{c}(i\omega)$ and $\Delta \Sigma_{f}(i\omega)$ are related
with non-Fermi liquid physics of such heavy fermions.

Static contributions of bosons determine the formation of the
heavy-fermion band, given by \bqa && \Phi_{c}(k,i\omega) = -
\frac{T V^{2} \rho_{b}}{N} G_{\psi}(0,0) G_{f}(k,i\omega) = -
\frac{V^{2} \rho_{b}}{N} \frac{T}{|\mu_{eff}^{\psi}|}
G_{f}(k,i\omega) \approx - \frac{V^{2} \rho_{b}}{N}
\frac{T}{|\mu_{eff}^{\psi}|} g_{f}(k,i\omega) , \nn &&
\Phi_{f}(k,i\omega) = - \frac{T V^{2} \rho_{b}}{N} G_{\psi}(0,0)
G_{c}(k,i\omega) = - \frac{V^{2} \rho_{b}}{N}
\frac{T}{|\mu_{eff}^{\psi}|} G_{c}(k,i\omega) \approx -
\frac{V^{2} \rho_{b}}{N} \frac{T}{|\mu_{eff}^{\psi}|}
g_{c}(k,i\omega) ,  \eqa where \bqa && \mu_{eff}^{\psi} =
\mu_{\psi} - \rho_{b} \lambda - \Sigma_{\psi}(0,0) \eqa is an
effective chemical potential, essential for coherence. When it
touches zero, $\frac{T}{|\mu_{eff}^{\psi}|}$ should be replaced
with $|\langle \psi \rangle|^{2}$. \bqa && g_{c}(k,i\omega) =
\frac{1}{i\omega + \mu_{c} - \frac{k^{2}}{2m_{c}}} , ~~~~~
g_{f}(k,i\omega) = \frac{1}{i\omega + \mu_{c} - \epsilon_{f} -
\lambda  - \frac{k^{2}}{2m_{f}}} \eqa are bare Green's functions.

Quantum fluctuations of bosons give rise to non-Fermi liquid
self-energy corrections of such heavy fermions \bqa && \Delta
\Sigma_{c}(i\omega) = - \frac{V^{2} \rho_{b}}{N} \frac{1}{\beta}
\sum_{i\Omega \not= 0} \int_{q \not= 0} \frac{d^{3}
q}{(2\pi)^{3}}G_{\psi}(q,i\Omega) G_{f}(k - q,i\omega - i\Omega) ,
\nn && \Delta \Sigma_{f}(i\omega) = - \frac{V^{2} \rho_{b}}{N}
\frac{1}{\beta} \sum_{i\Omega \not= 0} \int_{q \not= 0}
\frac{d^{3} q}{(2\pi)^{3}} G_{\psi}(q,i\Omega) G_{c}(k + q,i\omega
+ i\Omega) - \frac{1}{\beta} \sum_{i\Omega \not= 0} \int_{q \not=
0} \frac{d^{3} q}{(2\pi)^{3}} F(k,q) D (q,i\Omega)
G_{f}(k,i\omega) , \nn \eqa where the static component of the
$\psi$ propagator should not be taken into account.

For convenience, we also divide the $\psi$ self-energy as follows
\bqa && \Sigma_{\psi}(q,i\Omega) = \Sigma_{\psi}(0,0) + \Delta
\Sigma_{\psi}(q,i\Omega) , \eqa where the static contribution is
introduced into the effective chemical potential, given by \bqa &&
\Sigma_{\psi}(0,0) = V^{2}\rho_{b} \frac{1}{\beta} \sum_{i\omega}
\int \frac{d^{3} k}{(2\pi)^{3}} G_{f}(k,i\omega) G_{c}(k,i\omega)
\approx - \frac{V^{2} \rho_{b}}{2\pi^{2}} \frac{\mathcal{K}_{F}^{c
2}}{\mathcal{V}_{F}^{c}} , \eqa and the dynamic part results in
Landau damping, given by \bqa && \Delta \Sigma_{\psi}(q,i\Omega) =
V^{2}\rho_{b} \frac{1}{\beta} \sum_{i\omega} \int \frac{d^{3}
k}{(2\pi)^{3}} G_{f}(k,i\omega) G_{c}(k+q,i\omega+i\Omega) +
\rho_{b}^{2} \frac{1}{\beta} \sum_{i\nu} \int \frac{d^{3}
l}{(2\pi)^{3}} B(q,l) D (l,i\nu) G_{\psi}(q+l,i\Omega+i\nu) . \nn
\eqa In the next subsection we evaluate this dynamic contribution,
where the self-energy correction from gauge fluctuations (the
second term) will not be taken into account. That contribution is
irrelevant because the $\psi$ dynamics is described by $z = 3$,
which implies that the effective theory for the $\psi$ dynamics
lies above the upper critical dimension, resulting in the
mean-field-like dynamics. See our discussion on this issue in the
last section.

\subsection{To calculate the $\psi$ self-energy}

In order to cover both the quantum critical regime and the
incoherent heavy-fermion phase, we introduce the following fermion
Green's functions \bqa && G_{f}(i\omega,k) =
\frac{1}{i\omega-v_{F}^{f}(k-k_{F}^{f})-\frac{T
V^{2}\rho_{b}/|\mu_{eff}^{\psi}|} {i\omega-v_{F}^{c}(k-k_{F}^{c})}
- \Delta \Sigma_{f}(i\omega)} , \nn && G_{c}(i\omega,k) =
\frac{1}{i\omega-v_{F}^{c}(k-k_{F}^{c})-\frac{T
V^{2}\rho_{b}/|\mu_{eff}^{\psi}|} {i\omega-v_{F}^{f}(k-k_{F}^{f})}
- \Delta \Sigma_{c}(i\omega)} . \eqa $\Delta
\Sigma_{f(c)}(i\omega)$ is the spinon (electron) self-energy due
to inelastic scattering with quantum phase fluctuations, where the
heavy-fermion contribution of $\Phi_{f(c)}(k,i\omega)$ is
expressed explicitly. We emphasize that there is
$T/|\mu_{eff}^{\psi}|$, renormalizing $\rho_{b}$, which reduces
the strength of hybridization due to incoherence. We have
linearized each bare dispersion, where $v_{F}^{f(c)}$ is the Fermi
velocity and $k_{F}^{f(c)}$ is the Fermi momentum.

Neglecting non-Fermi liquid parts of self-energy corrections for
the time being, we can express Eq. (B9) as follows \bqa &&
G_{f/c}(\omega,k)=\frac{Z_{f/c}}{\omega-\mathcal{V}^{f/c}_{F}(k-\mathcal{K}_{F}^{f/c})}
, \eqa where $\mathcal{V}^{f/c}_{F}$ and $\mathcal{K}_{F}^{f/c}$
are renormalized Fermi velocity and renormalized Fermi momentum,
respectively, and $Z_{f/c}$ is the wave-function renormalization
function. They are given by \bqa && \mathcal{K}_{F}^{f} =
\mathcal{K}_{F}^{c} = \frac{1}{2}\left( k_{F}^{f}+k_{F}
^{c}\right) + \frac{1}{2}\sqrt{\left( k_{F}^{f}-k_{F}^{c}\right)
^{2}+4\frac{TV^{2}\rho_{b}/|\mu_{eff}^{\psi}|}{v_{F}^{f}v_{F}^{c}}}
, \nn && \mathcal{V}_{F}^{f/c} = Z_{f/c} \left[ v_{F}^{f/c} +
\frac{(TV^{2}\rho_{b}/|\mu_{eff}^{\psi}|) v_{F}^{c/f}}{\left(
v_{F}^{c/f}\right)^{2}\left(
\mathcal{K}_{F}-k_{F}^{c/f}\right)^{2}}\right] , ~~~ Z_{f/c}^{-1}
= 1 + \frac{TV^{2}\rho_{b}/|\mu_{eff}^{\psi}|}{\left(
v_{F}^{c/f}\right)^{2} \left(
\mathcal{K}_{F}-k_{F}^{c/f}\right)^{2}} . \eqa

Inserting Eq. (B9) with Eq. (B10) into the $\psi$ self-energy, we
obtain the following expression \cite{KB_Indranil,KB_Pepin} \bqa
&& \Delta \Sigma_{\psi}(i\Omega,\mathbf{q}) =
V^{2}\rho_{b}\frac{1}{\beta}\sum
\limits_{i\omega}\int\frac{d^{3}k}{(2\pi)^{3}}G_{f}(i\omega,\mathbf{k}
)G_{c}(i\omega+i\Omega,\mathbf{k+q}) \nn && = V^{2}\rho_{b}
\frac{(\mathcal{K}_{F}^{f})^{2}}{4\pi^{2}\mathcal{V}_{F}^{f}\mathcal{V}_{F}^{c}}
\frac {1}{\alpha-1}\frac{1}{q} \Bigl\{ \Bigl( \alpha
i\Omega+\mathcal{V}_{F}^{f}(\mathcal{K}_{F}^{c}-\mathcal{K}_{F}^{f}
) \Bigr) \log \Bigl( \frac{\alpha i\Omega +
\mathcal{V}_{F}^{f}(\mathcal{K}_{F}^{c}-\mathcal{K}_{F}^{f})+\alpha
\mathcal{V}_{F}^{c} q}{\alpha
i\Omega+\mathcal{V}_{F}^{f}(\mathcal{K}_{F}^{c}-\mathcal{K}_{F}^{f})-\alpha
\mathcal{V}_{F}^{c}q} \Bigr) \nn && - \Bigl(
i\Omega+\mathcal{V}_{F}^{f}(\mathcal{K}_{F}^{c}-\mathcal{K}_{F}^{f})
\Bigr) \log\Bigl( \frac{i\Omega+\mathcal{V}_{F}^{f}
(\mathcal{K}_{F}^{c}-\mathcal{K}_{F}^{f})+\alpha
\mathcal{V}_{F}^{c}q}{i\Omega+\mathcal{V}_{F}^{f}(\mathcal{K}_{F}^{c}
- \mathcal{K}_{F}^{f}) - \alpha \mathcal{V}_{F}^{c}q}\Bigr) \nn &&
+\alpha \mathcal{V}_{F}^{c}q \log\Bigl( \frac{(\alpha i\Omega +
\mathcal{V}_{F}^{f}(\mathcal{K}_{F}^{c}-\mathcal{K}_{F}^{f}))^{2}-(\alpha
\mathcal{V}_{F}^{c}q)^{2}}{(i\Omega
+\mathcal{V}_{F}^{f}(\mathcal{K}_{F}^{c}-\mathcal{K}_{F}^{f}))^{2}-(\alpha
\mathcal{V}_{F}^{c}q)^{2}} \Bigr) \Bigr\} \eqa with
$\alpha=\frac{\mathcal{V}_{F}^{f}}{\mathcal{V}_{F}^{c}}$. Taking
$\alpha \rightarrow 1$ with $\mathcal{K}_{F}^{c} =
\mathcal{K}_{F}^{f}$ for the heavy-fermion band, this expression
is simplified as \bqa && \Delta \Sigma_{\psi}(i\Omega,\mathbf{q})
= V^{2}\rho_{b}
\frac{(\mathcal{K}_{F}^{f})^{2}}{4\pi^{2}\mathcal{V}_{F}^{f}\mathcal{V}_{F}^{c}}
\frac{1}{q} \Bigl\{ i\Omega \ln \frac{i\Omega +
\mathcal{V}_{F}^{c}q}{i\Omega - \mathcal{V}_{F}^{c}q} + i\Omega
\Bigl( \frac{\mathcal{V}_{F}^{c}q}{i\Omega + \mathcal{V}_{F}^{c}q}
+ \frac{\mathcal{V}_{F}^{c}q}{i\Omega - \mathcal{V}_{F}^{c}q}
\Bigr) + 2 \mathcal{V}_{F}^{c}q \frac{(i\Omega)^{2}}{(i\Omega)^{2}
- (\mathcal{V}_{F}^{c}q)^{2}} \Bigr\} . \eqa

If one expands Eq. (B12) in the limit of
$\frac{\mathcal{K}^{f}_{F}-\mathcal{K}^{c}_{F}}{q} \ll 1$, the
typical Landau damping form results \bqa &&
\Sigma_{\psi}(i\Omega,q) = \gamma \frac{|\Omega|}{q} , \eqa
originating from particle-hole excitations around the Fermi
surface. The damping coefficient is given by \bqa && \gamma =
V^{2} \rho_{b} \frac{(\mathcal{K}_{F}^{f})^{2}}{4\pi
\mathcal{V}_{F}^{f}\mathcal{V}_{F}^{c}} . \eqa Then, the $\psi$
propagator becomes \bqa && G_{\psi}(q,i\Omega) \approx -
\frac{1}{\gamma \frac{|\Omega|}{q} + \frac{\rho_{b}}{2m_{b}} q^{2}
- \mu_{eff}^{\psi} } , \eqa where $\mu_{eff}^{\psi} = \mu_{\psi} -
\rho_{b} \lambda - \Sigma_{\psi}(0,0)$ is an effective chemical
potential for phase fluctuations.

Inserting this boson propagator into self-energy equations for
fermions, one can find fermion self-energy corrections. Such
calculations have been performed in previous studies
\cite{KB_Indranil,KB_Pepin} when the boson dynamics is critical
and described by $z = 3$. Since the $\psi$ dynamics is also
characterized by $z = 3$ when $T > T_{1}$, as discussed in the
manuscript, the previous results are applied to the present
situation directly. Then, we obtain non-Fermi liquid
self-energies.

\section{To derive self-consistent equations for three order
parameters of $\rho_{b}$, $\lambda$, and $\mu_{\psi}$}

\subsection{General formulae}

One can find self-consistent equations for order parameters from
the Luttinger-Ward free energy functional. An essential merit of
this approach is that vertex corrections are naturally introduced
beyond the Schwinger-Dyson equation for an order parameter usually
identified with the mean-field equation.

Minimizing the free energy functional with respect to $\rho_{b}$,
we obtain \bqa && u_{b} \rho_{b} + \frac{N}{\beta} \sum_{i\omega}
\int \frac{d^{3} k}{(2\pi)^{3}} \frac{\partial
\Sigma_{c}(k,i\omega)}{\partial \rho_{b}} G_{c} (k,i\omega) +
\frac{N}{\beta} \sum_{i\omega} \int \frac{d^{3} k}{(2\pi)^{3}}
\frac{\partial \Sigma_{f}(k,i\omega)}{\partial \rho_{b}} G_{f}
(k,i\omega) \nn && + \frac{1}{\beta} \sum_{i\Omega} \int
\frac{d^{3} q}{(2\pi)^{3}} \Bigl( i\Omega - \frac{q^{2}}{2m_{b}} -
\lambda - \frac{\partial \Sigma_{\psi}(k,i\omega)}{\partial
\rho_{b}} \Bigr) G_{\psi} (q,i\Omega) + \frac{1}{\beta}
\sum_{i\Omega} \int \frac{d^{3} q}{(2\pi)^{3}} \frac{\partial
\Pi(q,i\Omega)}{\partial \rho_{b}} D (q,i\Omega) = 0 , \eqa where
the derivative for each self-energy implies each vertex
correction. In particular, we see that $\frac{\partial
\Sigma_{\psi}(k,i\omega)}{\partial \rho_{b}}$ is the vertex
correction, performed in the single impurity problem
\cite{Read_KI}. Such a contribution is expected to modify the
slave-boson mean-field equation for $\rho_{b}$ in principle.

Minimizing the free energy functional with respect to $\lambda$,
we obtain \bqa && - \frac{\lambda}{u_{b}} - N S + \frac{N}{\beta}
\sum_{i\omega} \int \frac{d^{3} k}{(2\pi)^{3}} \frac{\partial
\Sigma_{c}(k,i\omega)}{\partial \lambda} G_{c} (k,i\omega) +
\frac{N}{\beta} \sum_{i\omega} \int \frac{d^{3} k}{(2\pi)^{3}}
\Bigl( 1 + \frac{\partial \Sigma_{f}(k,i\omega)}{\partial \lambda}
\Bigr) G_{f} (k,i\omega) \nn && + \frac{1}{\beta} \sum_{i\Omega}
\int \frac{d^{3} q}{(2\pi)^{3}} \Bigl( - \frac{1}{u_{b}} i\Omega -
\rho_{b} - \frac{\partial \Sigma_{\psi}(k,i\omega)}{\partial
\lambda} \Bigr) G_{\psi} (q,i\Omega) + \frac{1}{\beta}
\sum_{i\Omega} \int \frac{d^{3} q}{(2\pi)^{3}} \frac{\partial
\Pi(q,i\Omega)}{\partial \lambda} D (q,i\Omega) = 0 , \eqa where
$\frac{\partial \Sigma_{\psi}(k,i\omega)}{\partial \lambda}$ is
the vertex correction beyond the slave-boson mean-field analysis.

In the same way we find an equation for $\mu_{\psi}$, given by
\bqa && 1 + \frac{N}{\beta} \sum_{i\omega} \int \frac{d^{3}
k}{(2\pi)^{3}} \frac{\partial \Sigma_{c}(k,i\omega)}{\partial
\mu_{\psi}} G_{c} (k,i\omega) + \frac{N}{\beta} \sum_{i\omega}
\int \frac{d^{3} k}{(2\pi)^{3}} \frac{\partial
\Sigma_{f}(k,i\omega)}{\partial \mu_{\psi}} G_{f} (k,i\omega) \nn
&& + \frac{1}{\beta} \sum_{i\Omega} \int \frac{d^{3}
q}{(2\pi)^{3}} \Bigl( 1 - \frac{\partial
\Sigma_{\psi}(k,i\omega)}{\partial \mu_{\psi}} \Bigr) G_{\psi}
(q,i\Omega) + \frac{1}{\beta} \sum_{i\Omega} \int \frac{d^{3}
q}{(2\pi)^{3}} \frac{\partial \Pi(q,i\Omega)}{\partial \mu_{\psi}}
D (q,i\Omega) = 0 , \eqa where $\frac{\partial
\Sigma_{c}(k,i\omega)}{\partial \mu_{\psi}}$ and $\frac{\partial
\Sigma_{f}(k,i\omega)}{\partial \mu_{\psi}}$ are identified with
vertex corrections.

\subsection{An equation for $\mu_{\psi}$}

We analyze the equation for $\mu_{\psi}$. Inserting both fermion
self-energies associated with the heavy-fermion band and boson
self-energy into Eq. (C3), we obtain \bqa && 1 - \frac{N}{\beta}
\sum_{i\omega} \int \frac{d^{3} k}{(2\pi)^{3}} \Bigl\{
\frac{\partial }{\partial \mu_{\psi}} \Bigl( \frac{V^{2}
\rho_{b}}{N} \frac{T}{|\mu_{eff}^{\psi}|} g_{f}(k,i\omega) \Bigr)
\Bigr\} G_{c} (k,i\omega) - \frac{N}{\beta} \sum_{i\omega} \int
\frac{d^{3} k}{(2\pi)^{3}} \Bigl\{ \frac{\partial }{\partial
\mu_{\psi}} \Bigl( \frac{V^{2} \rho_{b}}{N}
\frac{T}{|\mu_{eff}^{\psi}|} g_{c}(k,i\omega) \Bigr) \Bigr\} G_{f}
(k,i\omega) \nn && + \frac{1}{\beta} \sum_{i\Omega} \int
\frac{d^{3} q}{(2\pi)^{3}} \Bigl\{ 1 + \frac{\partial }{\partial
\mu_{\psi}} \Bigl( \frac{V^{2} \rho_{b}}{2\pi^{2}}
\frac{\mathcal{K}_{F}^{c 2}}{\mathcal{V}_{F}^{c}} \Bigr) -
\frac{\partial }{\partial \mu_{\psi}} \Bigl( \gamma
\frac{|\Omega|}{q} \Bigr) \Bigr\} G_{\psi} (q,i\Omega) = 0 . \eqa
We rearrange this equation as follows \bqa && 1 + \frac{1}{\beta}
\sum_{i\Omega} \int \frac{d^{3} q}{(2\pi)^{3}} G_{\psi}
(q,i\Omega) + \frac{1}{\beta} \sum_{i\Omega} \int \frac{d^{3}
q}{(2\pi)^{3}} \Bigl\{ \frac{\partial }{\partial \mu_{\psi}}
\Bigl( \frac{V^{2} \rho_{b}}{2\pi^{2}} \frac{\mathcal{K}_{F}^{c
2}}{\mathcal{V}_{F}^{c}} \Bigr) - \frac{\partial }{\partial
\mu_{\psi}} \Bigl( \gamma \frac{|\Omega|}{q} \Bigr) \Bigr\}
G_{\psi} (q,i\Omega) \nn && = \frac{N}{\beta} \sum_{i\omega} \int
\frac{d^{3} k}{(2\pi)^{3}} \Bigl\{ \frac{\partial }{\partial
\mu_{\psi}} \Bigl( \frac{V^{2} \rho_{b}}{N}
\frac{T}{|\mu_{eff}^{\psi}|} g_{f}(k,i\omega) \Bigr) \Bigr\} G_{c}
(k,i\omega) + \frac{N}{\beta} \sum_{i\omega} \int \frac{d^{3}
k}{(2\pi)^{3}} \Bigl\{ \frac{\partial }{\partial \mu_{\psi}}
\Bigl( \frac{V^{2} \rho_{b}}{N} \frac{T}{|\mu_{eff}^{\psi}|}
g_{c}(k,i\omega) \Bigr) \Bigr\} G_{f} (k,i\omega) . \nn   \eqa
This expression is quite interesting in the respect that the first
two terms in the left-hand-side correspond to the Schwinger-Dyson
equation resulting from $\langle |\psi|^{2} \rangle$ = 1 while
other contributions originate from vertex corrections.

Keeping $\mu_{\psi}$-derivative terms only when they depend on
$\mu_{\psi}$ explicitly as the lowest-order approximation, we
reach the following expression \bqa && 1 - \frac{T V^{2}
\rho_{b}}{\mu_{eff}^{\psi 2}} \Bigl\{ \frac{1}{\beta}
\sum_{i\omega} \int \frac{d^{3} k}{(2\pi)^{3}} g_{f}(k,i\omega)
G_{c} (k,i\omega) + \frac{1}{\beta} \sum_{i\omega} \int
\frac{d^{3} k}{(2\pi)^{3}} g_{c}(k,i\omega) G_{f} (k,i\omega)
\Bigr\} = - \frac{1}{\beta} \sum_{i\Omega} \int \frac{d^{3}
q}{(2\pi)^{3}} G_{\psi} (q,i\Omega) . \nn \eqa It is clear that
fermion contributions are related with vertex corrections. We will
see that this correction plays an important role for
self-consistency, which cancels other quantum corrections.

\subsection{An equation for $\rho_{b}$}

Inserting both fermion self-energies associated with the
heavy-fermion band and boson self-energy into Eq. (C1), we obtain
\bqa && u_{b} \rho_{b} - \frac{N}{\beta} \sum_{i\omega} \int
\frac{d^{3} k}{(2\pi)^{3}} \Bigl\{ \frac{\partial }{\partial
\rho_{b}} \Bigl( \frac{V^{2} \rho_{b}}{N}
\frac{T}{|\mu_{eff}^{\psi}|} g_{f}(k,i\omega) \Bigr) \Bigr\} G_{c}
(k,i\omega) - \frac{N}{\beta} \sum_{i\omega} \int \frac{d^{3}
k}{(2\pi)^{3}} \Bigl\{ \frac{\partial }{\partial \rho_{b}} \Bigl(
\frac{V^{2} \rho_{b}}{N} \frac{T}{|\mu_{eff}^{\psi}|}
g_{c}(k,i\omega) \Bigr) \Bigr\} G_{f} (k,i\omega) \nn && +
\frac{1}{\beta} \sum_{i\Omega} \int \frac{d^{3} q}{(2\pi)^{3}}
\Bigl\{ i\Omega - \frac{q^{2}}{2m_{b}} - \lambda + \frac{\partial
}{\partial \rho_{b}} \Bigl( \frac{V^{2} \rho_{b}}{2\pi^{2}}
\frac{\mathcal{K}_{F}^{c 2}}{\mathcal{V}_{F}^{c}} \Bigr) -
\frac{\partial }{\partial \rho_{b}} \Bigl( \gamma
\frac{|\Omega|}{q} \Bigr) \Bigr\} G_{\psi} (q,i\Omega) = 0 . \eqa

Performing derivatives for $\rho_{b}$, we reach the following
expression \bqa && u_{b} \rho_{b} - \frac{T
V^{2}}{|\mu_{eff}^{\psi}|} \Bigl\{ \frac{1}{\beta} \sum_{i\omega}
\int \frac{d^{3} k}{(2\pi)^{3}} g_{f}(k,i\omega) G_{c} (k,i\omega)
+ \frac{1}{\beta} \sum_{i\omega} \int \frac{d^{3} k}{(2\pi)^{3}}
g_{c}(k,i\omega) G_{f} (k,i\omega) \Bigr\} \nn && + \frac{T V^{2}
\rho_{b}}{\mu_{eff}^{\psi 2}} \Bigl( \lambda -
\frac{V^{2}}{2\pi^{2}} \frac{\mathcal{K}_{F}^{c
2}}{\mathcal{V}_{F}^{c}} \Bigr) \Bigl\{ \frac{1}{\beta}
\sum_{i\omega} \int \frac{d^{3} k}{(2\pi)^{3}} g_{f}(k,i\omega)
G_{c} (k,i\omega) + \frac{1}{\beta} \sum_{i\omega} \int
\frac{d^{3} k}{(2\pi)^{3}} g_{c}(k,i\omega) G_{f} (k,i\omega)
\Bigr\} \nn && - \lambda \frac{1}{\beta} \sum_{i\Omega} \int
\frac{d^{3} q}{(2\pi)^{3}} G_{\psi} (q,i\Omega) +
\frac{V^{2}}{2\pi^{2}} \frac{\mathcal{K}_{F}^{c
2}}{\mathcal{V}_{F}^{c}} \frac{1}{\beta} \sum_{i\Omega} \int
\frac{d^{3} q}{(2\pi)^{3}} G_{\psi} (q,i\Omega) = -
\frac{1}{\beta} \sum_{i\Omega} \int \frac{d^{3} q}{(2\pi)^{3}}
\Bigl( i\Omega - \frac{q^{2}}{2m_{b}} \Bigr) G_{\psi} (q,i\Omega)
. \eqa

Inserting Eq. (C6) into the above equation, we obtain \bqa &&
u_{b} \rho_{b} - \frac{T V^{2}}{|\mu_{eff}^{\psi}|} \Bigl\{
\frac{1}{\beta} \sum_{i\omega} \int \frac{d^{3} k}{(2\pi)^{3}}
g_{f}(k,i\omega) G_{c} (k,i\omega) + \frac{1}{\beta}
\sum_{i\omega} \int \frac{d^{3} k}{(2\pi)^{3}} g_{c}(k,i\omega)
G_{f} (k,i\omega) \Bigr\} \nn && + \frac{T V^{2}
\rho_{b}}{\mu_{eff}^{\psi 2}} \Bigl( \lambda -
\frac{V^{2}}{2\pi^{2}} \frac{\mathcal{K}_{F}^{c
2}}{\mathcal{V}_{F}^{c}} \Bigr) \Bigl\{ \frac{1}{\beta}
\sum_{i\omega} \int \frac{d^{3} k}{(2\pi)^{3}} g_{f}(k,i\omega)
G_{c} (k,i\omega) + \frac{1}{\beta} \sum_{i\omega} \int
\frac{d^{3} k}{(2\pi)^{3}} g_{c}(k,i\omega) G_{f} (k,i\omega)
\Bigr\} \nn && + \lambda \Bigl[ 1 - \frac{T V^{2}
\rho_{b}}{\mu_{eff}^{\psi 2}} \Bigl\{ \frac{1}{\beta}
\sum_{i\omega} \int \frac{d^{3} k}{(2\pi)^{3}} g_{f}(k,i\omega)
G_{c} (k,i\omega) + \frac{1}{\beta} \sum_{i\omega} \int
\frac{d^{3} k}{(2\pi)^{3}} g_{c}(k,i\omega) G_{f} (k,i\omega)
\Bigr\} \Bigr] \nn && - \frac{V^{2}}{2\pi^{2}}
\frac{\mathcal{K}_{F}^{c 2}}{\mathcal{V}_{F}^{c}} \Bigl[ 1 -
\frac{T V^{2} \rho_{b}}{\mu_{eff}^{\psi 2}} \Bigl\{
\frac{1}{\beta} \sum_{i\omega} \int \frac{d^{3} k}{(2\pi)^{3}}
g_{f}(k,i\omega) G_{c} (k,i\omega) + \frac{1}{\beta}
\sum_{i\omega} \int \frac{d^{3} k}{(2\pi)^{3}} g_{c}(k,i\omega)
G_{f} (k,i\omega) \Bigr\} \Bigr] \nn && = - \frac{1}{\beta}
\sum_{i\Omega} \int \frac{d^{3} q}{(2\pi)^{3}} \Bigl( i\Omega -
\frac{q^{2}}{2m_{b}} \Bigr) G_{\psi} (q,i\Omega) . \eqa
Surprisingly, quantum corrections in the $\psi$ sector cancels
those in the fermion part, simplifying the above expression as
follows \bqa && u_{b} \rho_{b} - \frac{T
V^{2}}{|\mu_{eff}^{\psi}|} \Bigl\{ \frac{1}{\beta} \sum_{i\omega}
\int \frac{d^{3} k}{(2\pi)^{3}} g_{f}(k,i\omega) G_{c} (k,i\omega)
+ \frac{1}{\beta} \sum_{i\omega} \int \frac{d^{3} k}{(2\pi)^{3}}
g_{c}(k,i\omega) G_{f} (k,i\omega) \Bigr\} + \lambda -
\frac{V^{2}}{2\pi^{2}} \frac{\mathcal{K}_{F}^{c
2}}{\mathcal{V}_{F}^{c}} \nn && = - \frac{1}{\beta} \sum_{i\Omega}
\int \frac{d^{3} q}{(2\pi)^{3}} \Bigl( i\Omega -
\frac{q^{2}}{2m_{b}} \Bigr) G_{\psi} (q,i\Omega) . \eqa This
cancellation confirms the validity of our approximation in Eq.
(C6).

Neglecting the right-hand-side because we approximate the gauge
propagator as Eq. (B16), where the linear time-derivative is not
introduced, we reach the following expression \bqa && \lambda -
\frac{V^{2}}{2\pi^{2}} \frac{\mathcal{K}_{F}^{c
2}}{\mathcal{V}_{F}^{c}} = \frac{T V^{2}}{|\mu_{eff}^{\psi}|}
\Bigl\{ \frac{1}{\beta} \sum_{i\omega} \int \frac{d^{3}
k}{(2\pi)^{3}} g_{f}(k,i\omega) G_{c} (k,i\omega) +
\frac{1}{\beta} \sum_{i\omega} \int \frac{d^{3} k}{(2\pi)^{3}}
g_{c}(k,i\omega) G_{f} (k,i\omega) \Bigr\} , \eqa essentially the
same structure as that of the slave-boson mean-field theory except
for the second term in the left-hand-side.

\subsection{An equation for $\lambda$}

Inserting both fermion self-energies associated with the
heavy-fermion band and boson self-energy into Eq. (C2), we obtain
\bqa && - \frac{\lambda}{u_{b}} - N S - \frac{N}{\beta}
\sum_{i\omega} \int \frac{d^{3} k}{(2\pi)^{3}} \Bigl\{
\frac{\partial }{\partial \lambda} \Bigl( \frac{V^{2} \rho_{b}}{N}
\frac{T}{|\mu_{eff}^{\psi}|} g_{f}(k,i\omega) \Bigr) \Bigr\} G_{c}
(k,i\omega) \nn && + \frac{N}{\beta} \sum_{i\omega} \int
\frac{d^{3} k}{(2\pi)^{3}} \Bigl\{ 1 - \frac{\partial }{\partial
\lambda} \Bigl( \frac{V^{2} \rho_{b}}{N}
\frac{T}{|\mu_{eff}^{\psi}|} g_{c}(k,i\omega) \Bigr) \Bigr\} G_{f}
(k,i\omega) \nn && + \frac{1}{\beta} \sum_{i\Omega} \int
\frac{d^{3} q}{(2\pi)^{3}} \Bigl\{ - \frac{1}{u_{b}} i\Omega -
\rho_{b} + \frac{\partial }{\partial \lambda} \Bigl( \frac{V^{2}
\rho_{b}}{2\pi^{2}} \frac{\mathcal{K}_{F}^{c
2}}{\mathcal{V}_{F}^{c}} \Bigr) - \frac{\partial }{\partial
\lambda} \Bigl( \gamma \frac{|\Omega|}{q} \Bigr) \Bigr\} G_{\psi}
(q,i\Omega) = 0 . \eqa

Performing $\lambda$-derivatives for terms depending on $\lambda$
explicitly as the lowest-order approximation, we obtain \bqa && -
\frac{\lambda}{u_{b}} - N S + \frac{T V^{2}
\rho_{b}^{2}}{\mu_{eff}^{\psi 2}} \Bigl\{ \frac{1}{\beta}
\sum_{i\omega} \int \frac{d^{3} k}{(2\pi)^{3}} g_{f}(k,i\omega)
G_{c} (k,i\omega) + \frac{1}{\beta} \sum_{i\omega} \int
\frac{d^{3} k}{(2\pi)^{3}} g_{c}(k,i\omega) G_{f} (k,i\omega)
\Bigr\} \nn && + \frac{N}{\beta} \sum_{i\omega} \int \frac{d^{3}
k}{(2\pi)^{3}} G_{f} (k,i\omega) + \rho_{b} \Bigl[ 1 - \frac{T
V^{2} \rho_{b}}{\mu_{eff}^{\psi 2}} \Bigl\{ \frac{1}{\beta}
\sum_{i\omega} \int \frac{d^{3} k}{(2\pi)^{3}} g_{f}(k,i\omega)
G_{c} (k,i\omega) + \frac{1}{\beta} \sum_{i\omega} \int
\frac{d^{3} k}{(2\pi)^{3}} g_{c}(k,i\omega) G_{f} (k,i\omega)
\Bigr\} \Bigr] = 0 . \nn \eqa Also, quantum corrections in the
$\psi$ sector cancels those in the fermion part, recovering the
constraint equation in the slave-boson mean-field analysis \bqa &&
\rho_{b} + \frac{N}{\beta} \sum_{i\omega} \int \frac{d^{3}
k}{(2\pi)^{3}} G_{f} (k,i\omega) = N S , \eqa when the first term
in the right-hand-side of Eq. (C13) is neglected. This treatment
is consistent with the $\psi$ Green's function, where the
$\lambda/u_{b}$ term in the linear time-derivative is not
considered.

\section{To solve self-consistent equations for order parameters}

Three coupled self-consistent equations are given by \bqa && 1 -
\frac{T V^{2} \rho_{b}}{\mu_{eff}^{\psi 2}} \Bigl\{
\frac{1}{\beta} \sum_{i\omega} \int \frac{d^{3} k}{(2\pi)^{3}}
g_{f}(k,i\omega) G_{c} (k,i\omega) + \frac{1}{\beta}
\sum_{i\omega} \int \frac{d^{3} k}{(2\pi)^{3}} g_{c}(k,i\omega)
G_{f} (k,i\omega) \Bigr\} = - \frac{1}{\beta} \sum_{i\Omega} \int
\frac{d^{3} q}{(2\pi)^{3}} G_{\psi} (q,i\Omega) , \nn && \lambda -
\frac{V^{2}}{2\pi^{2}} \frac{\mathcal{K}_{F}^{c
2}}{\mathcal{V}_{F}^{c}} = \frac{T V^{2}}{|\mu_{eff}^{\psi}|}
\Bigl\{ \frac{1}{\beta} \sum_{i\omega} \int \frac{d^{3}
k}{(2\pi)^{3}} g_{f}(k,i\omega) G_{c} (k,i\omega) +
\frac{1}{\beta} \sum_{i\omega} \int \frac{d^{3} k}{(2\pi)^{3}}
g_{c}(k,i\omega) G_{f} (k,i\omega) \Bigr\} , \nn && \rho_{b} +
\frac{N}{\beta} \sum_{i\omega} \int \frac{d^{3} k}{(2\pi)^{3}}
G_{f} (k,i\omega) = N S \eqa beyond the mean-field analysis, where
quantum corrections are introduced self-consistently.

Inserting both renormalized heavy-fermion Green's functions and
renormalized $\psi$ propagator into Eq. (D1), we reach the final
formulae for two energy scales \bqa && 1 -
\frac{\rho_{b}}{|\mu_{eff}^{\psi}|} \Bigl( \lambda -
\frac{V^{2}}{2\pi^{2}} \frac{\mathcal{K}_{F}^{c
2}}{\mathcal{V}_{F}^{c}} \Bigr) = \frac{1}{\beta} \sum_{i\Omega}
\int \frac{d^{3} q}{(2\pi)^{3}} \frac{1}{\gamma \frac{|\Omega|}{q}
+ \frac{\rho_{b}}{2m_{b}} q^{2} - \mu_{eff}^{\psi} } , \nn &&
\lambda - \frac{V^{2}}{2\pi^{2}} \frac{\mathcal{K}_{F}^{c
2}}{\mathcal{V}_{F}^{c}} = 2 \frac{T V^{2}}{|\mu_{eff}^{\psi}|}
\int\limits_{0}^{\infty} \frac{d k}{2 \pi^2} k^2 \frac{f(E_{+}(k))
- f(E_{-}(k))}{E_{+}(k)-E_{-}(k)}  , \nn && 1 = \rho_{b} + 2
\int\limits_{0}^{\infty} \frac{d k}{2 \pi^2} k^2 \Big[ f(E_{-}(k))
\frac{v^{c}_{F}(k-k^{c}_{F}) - E_{-}(k)}{E_{+}(k)-E_{-}(k)} +
f(E_{+}(k)) \frac{
E_{+}(k)-v^{c}_{F}(k-k^{c}_{F})}{E_{+}(k)-E_{-}(k)} \Big] , \eqa
where the effective chemical potential and the renormalized
heavy-fermion band are \bqa && - \mu_{eff}^{\psi} \approx -
\mu_{\psi} + \rho_{b} \lambda - \frac{V^{2} \rho_{b}}{2\pi^{2}}
\frac{\mathcal{K}_{F}^{c 2}}{\mathcal{V}_{F}^{c}}
%\ln \Bigl( \frac{\mathcal{V}_{F}^{c}}{\mathcal{V}_{F}^{f}} \Bigr)
, \nn && E_{\pm}(k) =
\frac{1}{2}(\varepsilon_{f}(k)+\varepsilon_{c}(k)) \pm
\frac{1}{2}\sqrt{(\varepsilon_{f}(k)-\varepsilon_{c}(k))^2+4 T V^2
\rho_b / |\mu_{eff}^{\psi}|} , \eqa respectively. The damping
coefficient is given by \bqa && \gamma = V^{2} \rho_{b}
\frac{(\mathcal{K}_{F}^{f})^{2}}{4\pi
\mathcal{V}_{F}^{f}\mathcal{V}_{F}^{c}} , \eqa where renormalized
Fermi momentum and renormalized velocity are \bqa &&
\mathcal{K}_{F}^{f} = \mathcal{K}_{F}^{c} = \frac{1}{2}\left(
k_{F}^{f}+k_{F} ^{c}\right) + \frac{1}{2}\sqrt{\left(
k_{F}^{f}-k_{F}^{c}\right)^{2}+4\frac{T
V^{2}\rho_{b}/|\mu_{eff}^{\psi}|}{v_{F}^{f}v_{F}^{c}}} , \nn &&
\mathcal{V}_{F}^{f/c} = Z_{f/c} \left[ v_{F}^{f/c} + \frac{(T
V^{2}\rho_{b}/|\mu_{eff}^{\psi}|) v_{F}^{c/f}}{\left(
v_{F}^{c/f}\right)^{2}\left(
\mathcal{K}_{F}^{f/c}-k_{F}^{c/f}\right)^{2}}\right] , ~~~
Z_{f/c}^{-1} = 1 + \frac{T
V^{2}\rho_{b}/|\mu_{eff}^{\psi}|}{\left( v_{F}^{c/f}\right)^{2}
\left( \mathcal{K}_{F}^{f/c}-k_{F}^{c/f}\right)^{2}} . \eqa
$f(\epsilon)$ is the Fermi-Dirac distribution function. Then, all
quantities are defined, where $m_{c}$, $m_{f}$, $\epsilon_{f}$,
and $\mu_{c}$ are only parameters to define our system.

When vertex corrections are neglected in these equations, we
obtain \bqa && 1 = \frac{1}{\beta} \sum_{i\Omega} \int \frac{d^{3}
q}{(2\pi)^{3}} \frac{1}{\gamma \frac{|\Omega|}{q} +
\frac{\rho_{b}}{2m_{b}} q^{2} - \mu_{eff}^{\psi} } , \nn &&
\lambda = 2 \frac{T V^{2}}{|\mu_{eff}^{\psi}|}
\int\limits_{0}^{\infty} \frac{d k}{2 \pi^2} k^2 \frac{f(E_{+}(k))
- f(E_{-}(k))}{E_{+}(k)-E_{-}(k)}  , \nn && 1 = \rho_{b} + 2
\int\limits_{0}^{\infty} \frac{d k}{2 \pi^2} k^2 \Big[ f(E_{-}(k))
\frac{v^{c}_{F}(k-k^{c}_{F}) - E_{-}(k)}{E_{+}(k)-E_{-}(k)} +
f(E_{+}(k)) \frac{
E_{+}(k)-v^{c}_{F}(k-k^{c}_{F})}{E_{+}(k)-E_{-}(k)} \Big] , \eqa
where the second and third equations recover the slave-boson
mean-field equations, replacing $T/|\mu_{eff}^{\psi}|$ with
$|\langle\psi\rangle|^{2}$ in the second equation.

One may consider that the first equation of the rotor constraint
incorporates quantum corrections fully self-consistently because
the $\psi$ self-energy correction is introduced. On the other
hand, the fermion self-energy for non-Fermi liquid physics is not
taken into account. In our opinion introduction of such
self-energy corrections will not change the present picture, in
spite of modifying it only quantitatively.

\section{Numerical analysis}

%\subsection{To find $T^{*}$ and $T_{FL}$}

The higher energy scale $T^{*}$ is determined from
$\rho_{b}(T^{*}) = 0$. Then, self-consistent equations are reduced
to \bqa && 1 = T^{*} \sum_{i\Omega} \int \frac{d^{3}
q}{(2\pi)^{3}} \frac{1}{\gamma^{*} \frac{|\Omega|}{q} -
\mu_{eff}^{\psi *} }
%+ \frac{T^{*}}{|\mu_{eff}^{\psi *}|}
, ~~~~~ \gamma^{*} = {V}^{2}
\frac{k_{F}^{f}k_{F}^{c}}{v_{F}^{f}v_{F}^{c}} , \nn && \lambda^{*}
- \frac{V^{2}}{2\pi^{2}} \frac{k_{F}^{c 2}}{v_{F}^{c}} = 2
\frac{T^{*} V^{2}}{|\mu_{eff}^{\psi *}|} \int\limits_{0}^{\infty}
\frac{d k}{2 \pi^2} k^2 \frac{f^{*}(\epsilon_{f}(k)) -
f^{*}(\epsilon_{c}(k))}{\epsilon_{f}(k)-\epsilon_{c}(k)} , \nn &&
1 = 2 \int\limits_{0}^{\infty} \frac{d k}{2 \pi^2} k^2
f^{*}(\epsilon_{f}(k)) , \eqa where the conduction band is
decoupled from the spinon band. Notice that the boson band becomes
flat, resulting in incoherence as long as $- \mu_{eff}^{\psi} >
0$.

Solving the third equation, we obtain $\lambda$ as a function of
$T^{*}$. Inserting the $\lambda$ into the second equation, we find
$T^{*}/|\mu_{eff}^{\psi *}|$ as a function of both $T^{*}$ and
$V$. Inserting this function into the first equation, we obtain an
equation, representing the relation between $T^{*}$ and $V$, where
the following cutoff scheme is used, \bqa && \int_{0}^{\infty} d q
\int_{-\infty}^{\infty} d \nu g(q,\nu) = \frac{1}{\Lambda_{q}}
\int_{0}^{\Lambda_{q}} d q \frac{1}{2\Lambda_{\nu}}
\int_{-\Lambda_{\nu}}^{\Lambda_{\nu}} d \nu g(q,\nu) . \eqa As a
result, we find the $T^{*}(V)$ line in the phase diagram of Fig.
1.
%
%, where the blue-triangle line results from $\Lambda_{q} =
%\Lambda_{\nu} = 10$ and the red-circle line comes from
%$\Lambda_{q} = \Lambda_{\nu} = 50$.
%

It is subtle to determine the Fermi-liquid temperature. It is
identified with the condensation temperature of $\psi$, thus given
by $\mu_{eff}^{\psi}(T_{FL}) = 0$. Since
$T_{FL}/|\mu_{eff}^{\psi}(T_{FL})|$ diverges, self-consistent
equations are not well defined. It is natural to replace
$T/|\mu_{eff}^{\psi}|$ with $|\langle\psi\rangle|^{2}$. Taking
$\mu_{eff}^{\psi} = 0$ with $\langle\psi\rangle = 0$, we reach the
following equations to determine $T_{FL}$, \bqa && 1 = T_{FL}
\sum_{i\Omega} \int \frac{d^{3} q}{(2\pi)^{3}} \frac{1}{\gamma
\frac{|\Omega|}{q} + \frac{\rho_{b}}{2m_{b}} q^{2}} , ~~~~~ \gamma
= V^{2} \rho_{b} \frac{(\mathcal{K}_{F}^{f})^{2}}{4\pi
\mathcal{V}_{F}^{f}\mathcal{V}_{F}^{c}} , \nn && \lambda_{FL} -
\frac{V^{2}}{2\pi^{2}} \frac{\mathcal{K}_{F}^{c
2}}{\mathcal{V}_{F}^{c}} = 2 V^{2} \int\limits_{0}^{\infty}
\frac{d k}{2 \pi^2} k^2 \frac{f_{FL}(E_{+} (k)) - f_{FL} (E_{-}
(k))}{E_{+} (k)-E_{-} (k)} , \nn && 1 = \rho_{b} + 2
\int\limits_{0}^{\infty} \frac{d k}{2 \pi^2} k^2 \Big[ f_{FL}
(E_{-} (k)) \frac{v^{c}_{F}(k-k^{c}_{F}) - E_{-} (k)}{E_{+}
(k)-E_{-} (k)} + f_{FL} (E_{+} (k)) \frac{ E_{+}
(k)-v^{c}_{F}(k-k^{c}_{F})}{E_{+} (k)-E_{-} (k)} \Big] , \eqa
where \bqa && E_{\pm}(k) =
\frac{1}{2}(\varepsilon_{f}(k)+\varepsilon_{c}(k)) \pm
\frac{1}{2}\sqrt{(\varepsilon_{f}(k)-\varepsilon_{c}(k))^2+4 V^2
\rho_b} , \nn && \mathcal{K}_{F}^{f} = \mathcal{K}_{F}^{c} =
\frac{1}{2}\left( k_{F}^{f}+k_{F} ^{c}\right) +
\frac{1}{2}\sqrt{\left( k_{F}^{f}-k_{F}^{c}\right)^{2}+4 \frac{
V^{2}\rho_{b} }{v_{F}^{f}v_{F}^{c}}} , \nn &&
\mathcal{V}_{F}^{f/c} = Z_{f/c} \left[ v_{F}^{f/c} + \frac{(
V^{2}\rho_{b} ) v_{F}^{c/f}}{\left( v_{F}^{c/f}\right)^{2}\left(
\mathcal{K}_{F}^{f/c}-k_{F}^{c/f}\right)^{2}}\right] , ~~~
Z_{f/c}^{-1} = 1 + \frac{ V^{2}\rho_{b} }{\left(
v_{F}^{c/f}\right)^{2} \left(
\mathcal{K}_{F}^{f/c}-k_{F}^{c/f}\right)^{2}} . \eqa One may
criticize that Eq. (E4) are nothing but those of the heavy-fermion
state. Since $T/|\mu_{eff}^{\psi}|$ is replaced with
$|\langle\psi\rangle|^{2}$, one can claim that $V^{2} \rho_{b}$
should be modified into $V^{2} \rho_{b} |\langle\psi\rangle|^{2}$.
However, this substitution gives rise to a serious problem. When
$\langle\psi\rangle$ vanishes at $T = T_{FL}$, the resulting state
does not have the heavy-fermion band. This conclusion is in
contrast with the existence of $T^{*}$, where the heavy-fermion
band results, but its coherence is not achieved yet. In this
respect we perform the above approximation, where we resort to the
``bare'' heavy-fermion band. Although to use this band structure
overestimates $T_{FL}$, we believe that the final conclusion will
not change at least qualitatively.

It is straightforward to solve Eq. (E3) because the last two
equations are decoupled from the first equation. The last two
equations are nothing but the slave-boson mean-field equations.
Solving these coupled equations, we obtain $\lambda_{FL}$ and
$\rho_{b}$ as a function of both $T_{FL}$ and $V$. Inserting these
functions into the first equation, we can determine $T_{FL}(V)$ in
Fig. 1.

\section{Remarks}

\subsection{On decomposition for the RKKY spin-exchange term}

One may criticize the way how to take the RKKY spin-exchange
interaction in terms of spinons from Eq. (1) to Eq. (2). A
systematic description is to introduce the Sp(2N) representation,
which allows us to construct the spin operator in the large-N case
[Ying Ran and Xiao-gang Wen, arXiv:cond-mat/0609620]. Decomposing
the spin-exchange interaction into both spin-singlet and
spin-triplet channels and performing the Hubbard-Stratonovich
transformation for both exchange hopping and pairing interactions
in each spin channel, one can find an effective Hamiltonian for
paramagnetic Mott insulating phases (or spin liquids) [Ryuichi
Shindou and Tsutomu Momoi, Phys. Rev. B 80, 064410 (2009)]. All
these possible mean-field phases for spin liquids can be
classified in terms of the so called projective symmetry group
[Xiao-Gang Wen, Phys. Rev. B 65, 165113 (2002)]. Performing the
classification to distinguish various spin liquid states, one
takes into account the Gutzwiller projection for such mean-field
phases or integrates over gauge fluctuations accurately, and finds
actual paramagnetic ground states in each parameter regime. This
is the full procedure along this description.

However, this full procedure gives rise to much complication in
the analysis. Instead, we usually resort to experiments, helping
us neglect several irrelevant interactions in the system, which
will be determined by the Fermi-surface structure basically. The
Kondo breakdown scenario [I. Paul, C. P\'epin, and M. R. Norman,
Phys. Rev. Lett. {\bf 98}, 026402 (2007); C. P\'epin, Phys. Rev.
Lett. {\bf 98}, 206401 (2007)] has been applied to quantum
critical physics in YbRh$_{2}$Si$_{2}$, and successfully explained
thermodynamics [K.-S. Kim, A. Benlagra, and C. P\'epin, Phys. Rev.
Lett. {\bf 101}, 246403 (2008)], both electrical and thermal
transport [K.-S. Kim and C. P\'epin, Phys. Rev. Lett. {\bf 102},
156404 (2009)], and the Seebeck coefficient [Ki-Seok Kim and C.
P\'epin, Phys. Rev. B {\bf 81}, 205108 (2010); K.-S. Kim and C.
P\'epin, Phys. Rev. B {\bf 83}, 073104 (2011)] although this
theory underestimates antiferromagnetic correlations. Recall that
the Kondo breakdown theory is based on the uniform spin-liquid
ansatz for the antiferromagnetic phase, where quantum critical
physics is described by Fermi surface (Kondo) fluctuations,
expected to be applicable to the finite-temperature regime at
least. This quantum critical physics is described by the
Eliashberg approximation, where quantum corrections from
conduction electrons, spinons, holons, and even gauge fluctuations
are incorporated self-consistently. Maybe, the renormalization
group analysis will make our understanding of the present subject
deepen. But, we try to justify our effective Hamiltonian within
the phenomenological background.

\subsection{On phase fluctuations in the single impurity problem}

It is necessary to review the 1/N correction for the Kondo problem
in more detail. In particular, one may criticize our description
on the linearization for the phase factor, i.e., $e^{i\theta_{b}}
\approx 1 + i \theta_{b}$ because this procedure seems to break
the gauge invariance, which may cause unphysical divergences. In
the ``Higgs" phase such phase fluctuations should be eaten by
gauge fluctuations, giving rise to the fact that all field
variables in the resulting effective Hamiltonian are given by
gauge singlets, where the remaining gauge fluctuations (in the
unitary gauge) are gapped, and do not affect the low energy
physics. But, this should be regarded as just one way to describe
such phase fluctuations.

Actually, the linearization was proposed in N. Read, J. Phys. C:
Solid State Phys. {\bf 18}, 2651 (1985). Although this seminal
paper focuses on fluctuation corrections (1/N) in the complex
boson representation instead of the angular or polar coordinate
representation, the angular representation is also discussed
(section 6). As well discussed in the manuscript, such quantum
corrections give rise to the $log$-divergence for the expectation
value of holon condensation in the complex boson representation,
which implies that the condensation does not appear, where the
expectation value vanishes as the cutoff goes to infinity.
Actually, the holon propagator exhibits the power-law dependence
in time at long time scales instead of a constant.

Before we turn to the angular representation, we would like to
emphasize that an important point is how quantum corrections are
incorporated self-consistently well, keeping the Ward identity or
conservation law in any representations. Even if the perturbation
is used in the angular representation, the final result will be
physically valid as long as the Ward identity is satisfied. The
linearization gives rise to new type of a vertex, resulting in the
additional $log$-divergent contribution to the spinon self-energy.
This $log$-divergence was argued to cancel the $log$-divergence in
the expectation value of the holon condensation, where the
holon-condensation amplitude becomes finite. This means that the
holon condensation (complex number) vanishes due to transverse
(phase) fluctuations while the holon-condensation amplitude (real
number) remains finite. In this respect our present work can be
regarded to generalize the single-impurity problem to the
impurity-lattice problem. Indeed, our methodology is to generalize
that of the Kondo problem [N. Read, J. Phys. C: Solid State Phys.
{\bf 18}, 2651 (1985)] to the lattice problem.

The reason why we took this way is that we want to describe the
crossover regime out of the Higgs phase, where strong phase
fluctuations make the unitary gauge useless because singular
configurations for phase fluctuations, for example, vortices,
should be taken into account in that description [F. S. Nogueira
and H. Kleinert, arXiv:cond-mat/0303485, the World Scientific
review volume "Order, Disorder, and Criticality", Edited by Y.
Holovatch]. In this case the unitary gauge is not an easy way to
describe the crossover regime. Instead, we resort to the
non-linear $\sigma$-model description, allowing us to introduce
higher order quantum corrections and to describe the crossover
regime.

\subsection{On the holon self-interaction term}

It is necessary to discuss the physical origin of the holon
self-interaction term, phenomenologically introduced in the study.
Before this discussion, we would like to mention that this term
does not play any important role in low energy critical dynamics
because the Landau damping term is mostly relevant, which results
from dynamics of fermions near the Fermi surface.

Microscopically, the self-interaction term can originate from both
the J-term, i.e., $J \sum_{ij} (\vec{S}_{i}\cdot\vec{S}_{j} -
\frac{1}{4} n_{i} n_{j})$, where the density-interaction term with
the density operator $n_{i}$ of localized electrons is not shown
explicitly in the present manuscript, and the dynamically
generated self-interaction term resulting from quantum
fluctuations of fermions, where integrating over fermions and
expanding the resulting logarithm up to the fourth order for holon
excitations give rise to such a term. The second effect is
discussed in I. Paul, C. P\'epin, M. R. Norman, Phys. Rev. B {\bf
78}, 035109 (2008); C. P\'epin, Phys. Rev. B {\bf 77}, 245129
(2008). Since this effect appears from the short-distance scale,
such physics is not universal, irrelevant at low energies.

\subsection{On $\mu_{\psi}$ fluctuations}

One may criticize our approximation to neglect $\mu_{\psi}$
fluctuations for the $\psi$ dynamics. Actually, such fluctuations
play an important role in the nonlinear $\sigma$ model description
of the XY-type model, where the boson dynamics is described by $z
= 1$, i.e., the relativistic dispersion [T. Senthil, Phys. Rev. B
{\bf 78}, 045109 (2008)]. However, our phase-fluctuation dynamics
is governed by $z = 3$ due to Fermi-surface fluctuations, as
emphasized before. As a result, the boson dynamics can be
understood within the ``mean-field" approximation, which neglects
$\mu_{\psi}$ fluctuations, because the effective field theory for
the boson dynamics lives above the upper critical dimension, i.e.,
$d + z > d_{c} = 4$. In the same way gauge fluctuations do not
affect or alter the dynamics of boson excitations although they
are responsible for anomalous self-energy corrections to spinon
excitations. Although the situation is not completely the same as
our case, we would like to refer to T. Senthil, Phys. Rev. B {\bf
78}, 045109 (2008) for irrelevance of gauge-fluctuation
corrections to the holon (phase) self-energy.

Even if one focuses on the case ($z = 1$) when $\mu_{\psi}$
fluctuations are relevant, such fluctuations cannot erase the
existence of the incoherent heavy-fermion phase. They can modify
the critical physics in the preformed heavy-fermion phase, where
some critical exponents may be changed.

\end{widetext}


\begin{thebibliography}{9}
\bibitem{One_Loop_RG} All critical exponents are derived in the
Eliashberg approximation, giving rise to the same result as the
one-loop renormalization group analysis, where the vertex
correction of the ladder type does not affect critical exponents.
\end{thebibliography}
\end{document}